\documentclass[a4paper]{article}

\usepackage[left=3cm,right=3cm,top=3cm,bottom=3cm]{geometry}
\usepackage{authblk}

\usepackage{amsmath}
\usepackage{amssymb}
\usepackage{amsfonts}
\usepackage{amsthm}
\usepackage{adjustbox}
\usepackage{caption}
\usepackage{graphicx}
\usepackage{hyperref}

\DeclareMathOperator{\varset}{\mathsf{X}}
\DeclareMathOperator{\ta}{\mathsf{a}}
\DeclareMathOperator{\tb}{\mathsf{b}}
\DeclareMathOperator{\tc}{\mathsf{c}}
\DeclareMathOperator{\td}{\mathsf{d}}

\DeclareMathOperator{\varsx}{\mathsf{x}}
\DeclareMathOperator{\varsy}{\mathsf{y}}
\DeclareMathOperator{\varsz}{\mathsf{z}}

\DeclareMathOperator{\lang}{\mathcal{L}}
\DeclareMathOperator{\pclass}{\mathsf{P}}
\DeclareMathOperator{\npclass}{\mathsf{NP}}
\DeclareMathOperator{\supp}{\mathsf{sp}}
\DeclareMathOperator{\Supp}{\mathsf{supp}}

\begin{document}

\title{Extending Shinohara's Algorithm for Computing Descriptive (Angluin-Style) Patterns to Subsequence Patterns\thanks{Original research published in Kleest-Meißner, Sattler, Schmid, Schweikardt, Weidlich, "Discovering Event Queries from Traces: Laying Foundations for Subsequence-Queries with Wildcards and Gap-Size Constraints", ICDT 2022}}

\author[1]{Markus L.\ Schmid}

\affil[1]{Humboldt-Universit\"at zu Berlin, Germany, \texttt{MLSchmid@MLSchmid.de}}

\date{}

\maketitle

\begin{abstract}
The introduction of pattern languages in the seminal work [Angluin, ``Finding Patterns Common to a Set of Strings'', JCSS 1980] has revived the classical model of inductive inference (learning in the limit, gold-style learning). In [Shinohara, ``Polynomial Time Inference of Pattern Languages and Its Application'', 7th IBM Symposium on Mathematical Foundations of Computer Science 1982] a simple and elegant algorithm has been introduced that, based on membership queries, computes a pattern that is descriptive for a given sample of input strings (and, consequently, can be employed in strategies for inductive inference).\par
In this paper, we give a brief survey of the recent work [Kleest-Meißner et al., ``Discovering Event Queries from Traces: Laying Foundations for Subsequence-Queries with Wildcards and Gap-Size Constraints'', ICDT 2022], where the classical concepts of Angluin-style (descriptive) patterns and the respective Shinohara's algorithm are extended to a query class with applications in complex event recognition -- a modern topic from databases.
\end{abstract}

\section{Angluin-Style Patterns}

We briefly recall the concept of Angluin's pattern languages from~\cite{Angluin1980}.\par
A \emph{pattern} is a string over $\Sigma \cup \varset$, where $\Sigma$ is a finite alphabet of \emph{terminal symbols} and $\varset$ is a countable set of \emph{variables}. We use symbols $\ta, \tb, \tc, \td, \ldots$ as terminals and $\varsx, \varsy, \varsz, \varsx_1, \varsx_2, \ldots, \varsy_1, \varsy_2, \ldots$ as variables.\par
Such a pattern, e.\,g., $\alpha = \varsx_1 \ta \tb \varsx_2 \varsx_1 \ta \varsx_3 \tb \tc \varsx_2$, represents a \emph{pattern language} $\lang(\alpha)$ that contains exactly the words over $\Sigma$ that can be obtained from $\alpha$ by uniformly replacing the variables by non-empty words over $\Sigma$. For example, $\tb \ta \ta \tb \tc \ta \tc \tb \ta \ta \tc \tb \tc \tc \ta \tc \in \lang(\alpha)$ via the mapping $\varsx_1 \mapsto \tb \ta$, $\varsx_2 \mapsto \tc \ta \tc$ and $\varsx_3 \mapsto \tc$. \par
More formally, any mapping $h : \varset \to \Sigma^+$ is a \emph{substitution}, and we consider its natural extension to $h : (\varset \cup \Sigma) \to \Sigma^+$ by letting $h$ be the identity on terminals of $\Sigma$, and we finally consider its natural extension to a morphism $(\Sigma \cup \varset)^+ \to \Sigma^+$. Then, for every pattern $\alpha \in (\Sigma \cup \varset)^+$, its pattern language is defined by $\lang(\alpha) = \{h(\alpha) \mid h : \varset \to \Sigma^+\}$.\par
Compared to other classical formal language classes, pattern languages can be defined in a very simple way without relying on an automaton model or some generative device like a grammar or expressions. On the other hand, their expressive power is somewhat orthogonal to the Chomsky hierarchy: most pattern languages are neither regular nor context-free (e.\,g., the language described by the pattern $\varsx \varsx \varsx$), and there are trivial context-free and even regular languages that cannot be expressed as a pattern language (e.\,g., every finite and non-singleton language is not a pattern language). \par
An important result by~\cite{Gold1967} demonstrates that if a language class contains all finite and at least one infinite language, then it cannot be learned in the framework of inductive inference. In particular, this means that no language class that extends regular languages is suitable for inductive inference; thus, Gold's result has been interpreted as a rather negative result for inductive inference. In this regard, the incapability of pattern languages to describe all finite languages is an asset that entails the possibility of inductive inference for this language class. Following this observation, characterisations for language classes learnable by inductive inference have been produced (see~\cite{Angluin1980_2}).

\section{Descriptive Patterns and Shinohara's Algorithm}\label{sec:knownResults}

For a given finite sample $S = \{w_1, w_2, \ldots, w_k\}$ of words over $\Sigma$, a pattern $\alpha$ is said to be \emph{descriptive} if $S \subseteq \lang(\alpha)$ and there is no other pattern $\beta$ with $S \subseteq \lang(\beta) \subsetneq \lang(\alpha)$. Intuitively, this concept describes the situation that $\alpha$ is a suitable descriptor of $S$, and there is no other pattern that describes $S$ in a better way with respect to the subset relation, i.\,e., $\lang(\alpha)$ contains $S$ and is inclusion minimal with respect to all pattern languages that contain $S$.\par
In~\cite{Shinohara1982}, a simple and elegant algorithm is presented that, given a sample $S$, computes a pattern that is descriptive for $S$. We now sketch \emph{Shinohara's algorithm}.\par
Let $w = a_1 a_2 \ldots a_m$ be a shortest word from $S$. We start with a most general pattern $\alpha_1 =  \varsx_1 \varsx_2 \ldots \varsx_m$ and move over this pattern from left to right, i.\,e., we consider the first position, then the second position and so on. In these $m$ steps of the algorithm, we \emph{refine} the current pattern $\alpha_i$ to a pattern $\alpha_{i + 1}$ by possibly manipulating position $i$ of $\alpha_i$. Assume that we reached step $i$ of the algorithm, i.\,e., the current pattern is $\alpha_i$ and we consider the $i^{\text{th}}$ position $\alpha_i[i] = \varsx_i$. We replace $\varsx_i$ by $a_i$ (i.\,e., $w$'s $i^{\text{th}}$ symbol) and check whether the thus modified pattern $\alpha_{i + 1}$ can still describe $S$, i.\,e., $S \subseteq \lang(\alpha_{i+1})$. If this test is successful, then we move on to the next position, if the test fails, we try to replace $\varsx_i$ by any of the variables that occur in the prefix $\alpha_i[1..i-1]$ (note that there might not be any variables in this prefix). Again, if any such replacement yields a pattern $\alpha_{i + 1}$ with $S \subseteq \lang(\alpha_{i+1})$, then we move on to the next position. If no modification is possible, i.\,e., replacing $x_i$ by $a_i$ or by any of the variables from $\alpha_i[1..i-1]$ yields a pattern that cannot describe all words of $S$, then we simply keep the original variable $\varsx_i$ at position $i$ and move on to the next position $i + 1$; this means that $\alpha_{i + 1} = \alpha_i$ (note that this also means that $\varsx_i$ has become one of the variables that occur in the prefix $\alpha_{i'}[1..i'-1]$ in further steps $i'$ of the algorithm, and therefore can be used to replace variable $x_{i'}$).\par
The correctness of this algorithm hinges on the fact that for patterns $\alpha$ and $\beta$ of the same size, the inclusion $\lang(\alpha) \subseteq \lang(\beta)$ is characterised by the existence of a morphism $h : (\Sigma \cup \varset)^* \to (\Sigma \cup \varset)^*$ with $h(\beta) = \alpha$ (note that in the general case, the inclusion problem for pattern languages is undecidable; see~\cite{JiangEtAl1995,FreydenbergerReidenbach2010}). This means that the patterns $\alpha_1, \alpha_2, \ldots, \alpha_{m+1}$ of the $m$ steps of Shinohara's algorithm are in fact refinements in the sense that $\lang(\alpha_1) \supseteq \lang(\alpha_2) \supseteq \ldots \supseteq \lang(\alpha_{m+1})$. But why is $\alpha_{m+1}$ descriptive? If there were some $\beta$ with $S \subseteq \lang(\beta) \subsetneq \lang(\alpha_{m+1})$, then $|\beta| = |\alpha_{m+1}|$ (since $|\alpha_{m+1}| < |\beta|$ implies $w \notin \lang(\beta)$, and $|\beta| < |\alpha_{m+1}|$ implies $\lang(\beta) \not\subseteq \lang(\alpha_{m+1})$). Now assume that $\alpha_{m+1}[1..\ell] = \beta[1..\ell]$ and $\alpha_{m+1}[\ell+1] \neq \beta[\ell + 1]$. Due to the characterisation of containment by the existence of a morphism mapping one pattern to the other, we know that $\lang(\beta) \subsetneq \lang(\alpha_{m+1})$ implies that $\alpha_{m+1}[\ell + 1]$ is a variable $\varsx_{q}$, and it can even be assumed that $q = \ell + 1$ (i.\,e., position $\ell + 1$ is an original variable that is not replaced by Shinohara's algorithm). Hence, in step $(\ell + 1)$ of the algorithm, there is no suitable replacement for variable $\varsx_{\ell + 1}$. This, however, is a contradiction, since $\beta$'s $(\ell + 1)^{\text{th}}$ symbol (which, by assumption, is different from $\varsx_{\ell + 1}$) is in fact a suitable replacement in the $(\ell + 1)^{\text{th}}$ step of the algorithm. \par
Obviously, we can always find a descriptive pattern by an exhaustive search of the finite (yet exponentially large) set of all patterns that describe $S$. In this regard, an intuitive point of view of Shinohara's algorithm is that it traverses this search space rather efficiently: it starts with a most general pattern $\alpha_1 = \varsx_1 \varsx_2 \ldots \varsx_m$ (where $m = \min\{|w| \mid w \in S\}$) and then in only $m$ steps it moves along a chain of patterns that are getting more and more specific until it stops at a descriptive pattern. \par
One important aspect is to be discussed in more detail. Shinohara's algorithm must solve queries of the form ``$S \subseteq \lang(\alpha_i)$'', i.\,e., \emph{membership queries}, which, in general, constitutes an $\npclass$-complete problem (note that the membership problem ``$w \in \lang(\alpha)$?'' is a general type of \emph{pattern matching problem} that appears in various different contexts and its complexity has been intensively studied over the last decade, see~\cite{FernauSchmid2015,FernauEtAl2016,ReidenbachSchmid2014,FernauEtAl2020,DayEtAl2017} and the survey~\cite{ManeaSchmid2019}). In particular, there are many classes of patterns for which the membership problem can be solved efficiently, and for any such class $\mathfrak{P}$, we can restrict Shinohara's algorithm to produce \emph{$\mathfrak{P}$-descriptive} patterns $\alpha$, i.\,e., $\alpha \in \mathfrak{P}$, $S \subseteq \lang(\alpha)$ and there is no other pattern $\beta \in \mathfrak{P}$ with $S \subseteq \lang(\beta) \subsetneq \lang(\alpha)$. This only requires to add to the check ``$S \subseteq \lang(\alpha_{i + 1})$?'' an additional check ``$\alpha_{i + 1} \in \mathfrak{P}$?''. This small observation is very important, since it means that for any $\mathfrak{P}$ with efficient membership problem, Shinohara's algorithm is efficient (this observation has been used in Shinohara's original paper with respect to the class of so-called \emph{regular} and \emph{non-cross} patterns). \par
Consequently, in terms of complexity, Shinohara's algorithm is a way to reduce the computation of $\mathfrak{P}$-descriptive patterns to the computation of the membership problem. Is there any other way to compute descriptive patterns that \emph{does not rely} on the intractable task of checking membership? As shown in~\cite{FernauEtAl2018}, the answer is no. More precisely, any algorithm that computes $\mathfrak{P}$-descriptive patterns can be used for solving the membership problem for the class $\mathfrak{P}$. Consequently, for any class $\mathfrak{P}$ of patterns, we can efficiently solve the membership problem if and only if we can efficiently compute $\mathfrak{P}$-descriptive patterns (and this statement is constructive).

\section{Subsequence Patterns With Length Constraints}

The results outlined in Section~\ref{sec:knownResults} are classical in the field of inductive inference (note that we do not discuss here how computing descriptive patterns entails strategies for inferring pattern languages in the model of inductive inference; the interested reader is referred to~\cite{Shinohara1982,Angluin1980_2}). However, Angluin-style patterns as language descriptors in combination with the descriptiveness property (as a measure for how well patterns describe finite samples) and Shinohara's algorithm provide a rather general algorithmic framework which can be extended to other interesting learning or inference tasks. This has been demonstrated in~\cite{KleestMeissnerEtAl2022} with respect to so-called \emph{subsequence patterns} (with \emph{gap size constraints}) that can be applied in the context of complex event recognition. We shall next outline this setting.\par
We interpret Angluin-style patterns, e.\,g., $\alpha = \ta \varsx_1 \tb \varsx_2 \varsx_1 \ta \tc \varsx_2$, as \emph{subsequence patterns}, i.\,e., $\alpha$ matches a word $w$ if there is a substitution $h$ for the variables such that $h(\alpha)$ is a \emph{subsequence} (also called \emph{scattered factor} or \emph{subword} in the literature) of $w$. Moreover, we restrict the substitutions to be of the form $h : \varset \to \Sigma$, i.\,e., the variables range over single symbols instead of words from $\Sigma^+$. Finally, we also add \emph{gap size constraints}, which is a tuple $C_{\alpha} = (C_1, C_2, \ldots, C_{|\alpha| - 1})$ of lower and upper bounds $C_i = (c^-_i, c^+_i) \in \mathbb{N}^2$. Such gap size constraints $C_{\alpha}$ restrict the size of the gaps induced by the subsequence embedding of $h(\alpha)$ into $w$.\footnote{The model in~\cite{KleestMeissnerEtAl2022} also contains a \emph{global window size}, which upper bounds the total area the subsequence is mapped to; for simplicity, we ignore this constraint in this survey.} For example, if $\alpha = \ta \varsx_1 \tb \varsx_1$ and $C_{\alpha} = ((1, 3), (4, 4), (2, 3))$, then $\alpha$ matches a string $w$ only if we can substitute $\varsx_1$ by some symbol $\sigma \in \Sigma$, such that the resulting string can be embedded as a subsequence into $w$ in such a way that the first symbol $\ta$ and the second symbol $\sigma$ are mapped to positions $i_1$ and $i_2$ of $w$ with $1 \leq (i_2 - i_1 - 1) \leq 3$, the third symbol $\tb$ is mapped to a position $i_3$ with $4 \leq (i_3 - i_2 - 1) \leq 4$, and the fourth symbol $\sigma$ is mapped to a position $i_4$ with $2 \leq (i_4 - i_3 - 1) \leq 3$. For example, $\alpha$ matches $w = \ta \ta \tb \ta \tc \ta \tb \tc \tb \tb \ta \tc \tc$ in the following way:
\begin{align*}
&\alpha = & && &\ta& && && &\varsx_1& && && && && &\tb& && && &\varsx_1& &&\\
&& && &\downarrow& && && &\downarrow& && && && && &\downarrow& && && &\downarrow& &&\\
&w = &&\ta& &\ta& &\tb& &\ta& &\tc& &\ta& &\tb& &\tc& &\tb& &\tb& &\ta& &\tc& &\tc&
\end{align*}
Note that $\varsx_1$ is mapped to $\tc$, and the three gaps induced by the subsequence embedding are $2$, $4$ and $2$, respectively. Due to symbols $\ta$ and $\tb$ in $\alpha$, and the first two gap size constraints $(1, 3)$ and $(4, 4)$, any suitable embedding must map $\alpha$'s first and third symbol to an occurrence of $\ta$ and $\tb$ in $w$ with at least $6$ symbols in between. This means that $\tb$ of $\alpha$ must be mapped to the $9^{\text{th}}$ or $10^{\text{th}}$ position of $w$. However, choosing the $9^{\text{th}}$ position immediately implies that the first occurrence of $\varsx_1$ is mapped to the occurrence of $\ta$ on position $4$ of $w$ (due to the gap size constraint $(4, 4)$), which makes it impossible to map the second occurrence $\varsx_1$ to an occurrence of $\ta$ in such a way that the gap constraint $(2, 3)$ is satisfied. Consequently, the depicted embedding is the only possible one.\par
In the following section, we shall outline that the framework of descriptive patterns and Shinohara's algorithm can be extended to this setting of subsequence patterns with gap size constraints. Let us now discuss some properties of this modification.\par
For a subsequence pattern $\alpha$ with gap size constraints $C_{\alpha}$, we can also denote by $\lang(\alpha, C_{\alpha})$ the set of all words that are matched by $\alpha$ according to the semantics explained above. Unlike for classical Angluin-style patterns, every $\lang(\alpha, C_{\alpha})$ is a regular language, i.\,e., the language class associated with subsequence patterns with gap size constraints is a subclass of the regular languages. However, it is again not possible to describe finite and non-singleton languages. \par
It is shown in~\cite{KleestMeissnerEtAl2022} that the complexity of matching subsequence patterns is similar to that of classical Angluin-style patterns: In general, the problem is $\npclass$-complete, even if $|\Sigma| = 2$ and the gap size constraints are only $(0, 1)$; moreover, fixed-parameter tractability is not possible with respect to the rather strong parameter of the complete pattern size $|\alpha|$ (see~\cite{KleestMeissnerEtAl2022} for some fixed-parameter tractable variants).

\section{Shinohara's Algorithm For Subsequence Patterns}

Due to the gap size constraints, it is not straightforward to extend the concept of descriptiveness to subsequence patterns. In order to apply a variant of Shinohara's algorithm, we want to mimic the situation that we have for Angluin-style patterns, namely that inclusion $\lang(\alpha) \subseteq \lang(\beta)$ for patterns $\alpha$ and $\beta$ of equal length is characterised by a substitution that maps $\beta$ to $\alpha$. For subsequence patterns, this is also the case if in addition to the length of the patterns, also their tuples of gap size constraints are the same, i.\,e., for subsequence patterns $\alpha$ and $\beta$ with $|\alpha| = |\beta| = \ell$ and an $(\ell-1)$-tuple $C$ of gap size constraints, we have $\lang(\alpha, C) \subseteq \lang(\beta, C)$ if and only if there is a substitution that maps $\beta$ to $\alpha$. For the concept of descriptiveness, this requires the (rather strong) restriction that we have to fix the tuple of gap size constraints and therefore also the length of the subsequence patterns. On the other hand, we can generalise the concept of descriptiveness by requiring that a sufficiently large part of $S$ can be described by $\lang(\alpha, C)$, instead of requiring $S \subseteq \lang(\alpha, C)$. More precisely, we define $\Supp((\alpha, C), S) =  \frac{|\{w \in S \mid w \in \lang(\alpha, C)\}|}{|S|}$ and then require $\Supp((\alpha,C),S) \geq \supp$ for a given \emph{support threshold} $0 < \supp \leq 1$.\par
More formally, the concept of descriptiveness for subsequence patterns is defined as follows. Let $S$ be a finite sample and let $\mathfrak{P}$ be a set of subsequence patterns. For a fixed $\ell \in \mathbb{N}$, an $(\ell - 1)$-tuple $C$ of gap size constraints and a support threshold $\supp$ with $0 < \supp \leq 1$, a subsequence pattern $(\alpha, C)$ is \emph{descriptive for $S$ w.r.t. $(\mathfrak{P}, \supp, \ell, C)$} if $(\alpha, C) \in \mathfrak{P}$, $\Supp((\alpha,C),S) \geq \supp$ and
there is no subsequence pattern $(\beta, C) \in \mathfrak{P}$ with $\Supp((\beta, C), S) \geq \supp$ and $\lang(\beta, C) \subsetneq \lang(\alpha, C)$.\par
With this definition of descriptiveness of subsequence patterns, we can devise a variant of Shinohara's algorithm that, for an $(\ell-1)$-tuple $C$ of gap constraints, a sample $S$ and a support threshold $\supp$, computes a subsequence pattern that is descriptive for $S$ w.r.t. $(\mathfrak{P}, \supp, \ell, C)$. However, in comparison to Shinohara's original algorithm for Angluin-style patterns, we formulate the algorithm in a more general way (also motivated by the application in complex event recognition).\par
We start with the most general subsequence pattern $\alpha = x_1 x_2 \ldots x_{\ell}$ (note that the corresponding $(\ell-1)$-tuple $C$ of gap size constraints is part of the input). Then we visit all positions of $\alpha$ like in the original Shinohara's algorithm, but we can do this in any order, not necessarily from left to right. For every visited positions $j$, we consider all substitution of $x_j$ by a terminal symbol or some of the variables of already visited positions, and this can be done in any order. As soon as one of these substitutions is successful (in the sense that the current $(\alpha, C)$ is in $\mathfrak{P}$ and sufficiently covers the sample $S$ according to the required input support $\supp$), we keep this substitution and move on. If all substitutions fail, we keep variable $x_j$ at position $j$, which means that from now on $x_j$ is available as one of the variables further positions can be replaced with. \par
Analogously as in the case for Angluin-style patterns, the correctness of this algorithm follows from the characterisation of the inclusion by substitutions. The parameterisation by a class $\mathfrak{P}$ (note that just like for the original Shinohara's algorithm (see~\cite{FernauEtAl2018}), the class $\mathfrak{P}$ must satisfy some mild closure properties) is vital, since it allows to apply the algorithm with respect to classes of subsequence patterns with tractable matching problem, which makes the algorithm run in polynomial time. The extension to a support threshold is also possible for Angluin-style patterns, but especially interesting in the application context of~\cite{KleestMeissnerEtAl2022}. A particularly interesting observation is that the algorithm can also be started with \emph{any} length-$\ell$ subsequence pattern $(\alpha, C)$ (instead of $\alpha = x_1 x_2 \ldots x_{\ell}$), and then the algorithm computes a descriptive pattern $\alpha'$ that is more specific than $\alpha$ in the sense that $\lang(\alpha', C) \subseteq \lang(\alpha, C)$. Moreover, if the algorithm does not change the input pattern $\alpha$, i.\,e., the output is $\alpha'$ with $\alpha = \alpha'$, then the input pattern $\alpha'$ must already be descriptive. Thus, the algorithm can also be used for checking whether a given pattern is descriptive.\par
Another property of the original setting is that also for subsequence patterns we can show that for any class $\mathfrak{P}$ of subsequence patterns the matching problem can be solved in polynomial time if and only if descriptive subsequence patterns for $\mathfrak{P}$ can be computed in polynomial time (subject to the assumption $\pclass \neq \npclass$). Just like for Angluin-style patterns, this provides a strong motivation for algorithms that compute descriptive patterns based on solving the matching problem, since this is inherently done by any algorithm for computing descriptive patterns.

\subsection{Application to Complex Event Recognition}

In the area of databases, \emph{event stream processing} (and \emph{complex event recognition}) has recently emerged as a computational paradigm that is based on the continuous evaluation of queries over event streams. Several individual events of some system are given in form of a stream (e.\,g., ordered by their time stamp), and the task is to infer or evaluate patterns over such event streams. A typical feature of this setting is that patterns of interest of such even streams are formulated as subsequences of the event streams that have certain additional properties. The approach of~\cite{KleestMeissnerEtAl2022} is formalised by the subsequence patterns as described above, and the recognition task is formulated by computing descriptive subsequence patterns for a sample of event streams. We refer to the introduction of~\cite{KleestMeissnerEtAl2022} for more details on how subsequence patterns can describe situations of interest in event streams. \par

\end{document}